# High resolution time- and angle-resolved photoemission spectroscopy with 11 eV laser pulses

Changmin Lee[1,5†], Timm Rohwer[1,6†], Edbert J. Sie[1,7†], Alfred Zong[1], Edoardo Baldini[1], Joshua Straquadine[2,3,4], Philip Walmsley[2,3,4], Dillon Gardner[1], Young S. Lee[1,3,4], Ian R. Fisher[2,3,4], and Nuh Gedik[1*]

[1] Department of Physics, Massachusetts Institute of Technology, Cambridge, MA, USA

[2] Geballe Laboratory for Advanced Materials, Stanford University, Stanford, CA, USA

[3] Department of Applied Physics, Stanford, CA, USA

[4] SIMES, SLAC National Accelerator Laboratory, Menlo Park, CA, USA

[5] Present address: Division of Materials Science, Lawrence Berkeley National Laboratory, Berkeley, CA, USA

[6] Present address: Center for Free-Electron Laser Science, Deutsches Elektronen Synchrotron, Hamburg, Germany

[7] Present address: Geballe Laboratory for Advanced Materials, Stanford University, Stanford, CA, USA

† These authors equally contributed to this work

* gedik@mit.edu

**Performing time and angle resolved photoemission spectroscopy (tr-ARPES) at high momenta necessitates extreme ultraviolet laser pulses, which are typically produced via high harmonic generation (HHG). Despite recent advances, HHG-based setups still require large pulse energies (hundreds of μJ to mJ) and their energy resolution is limited to tens of meV. Here, we present a novel 11 eV tr-ARPES setup that generates a flux of $5 \times 10^{10}$ photons/s and achieves an unprecedented energy resolution of 16 meV. It can be operated at high repetition rates (up to 250 kHz) while using input pulse energies down to 3 μJ. We demonstrate these unique capabilities by simultaneously capturing the energy and momentum**

**resolved dynamics in two well-separated momentum space regions of a charge density wave material ErTe$_3$. This novel setup offers opportunity to study the non-equilibrium band structure of solids with exceptional energy and time resolutions at high repetition rates.**

**Introduction**

Angle-resolved photoemission (ARPES) has been at the forefront in the study of solids. Its capability of directly revealing the band structure has helped scientists to study the electronic properties of various systems ranging from metals and semiconductors to superconductors and topological materials. With the technological advance in femtosecond laser systems and nonlinear optics, time-resolved measurements of ARPES spectra in a tabletop setting are now becoming more common[1-3]. In a typical time-resolved ARPES (tr-ARPES) measurement, one laser pulse excites the system to higher energies, and a subsequent ultraviolet pulse ejects the photoelectrons from the solid. Tuning the time delay between the two pulses allows measurement of the non-equilibrium band structure, which often reveals unoccupied states in the conduction band[4,5], presence of collective modes such as phonons or charge density wave (CDW) amplitude modes[6], and creation of exotic non-equilibrium electronic states[7,8] that are not present in the absence of the laser pulse excitation.

As femtosecond lasers generally operate in the near-infrared regime, multiple nonlinear crystals (e.g. β-barium borate) are employed to generate ultraviolet laser pulses through frequency multiplication. With such crystals however, output wavelengths are capped to ~190 nm (6.5 eV) due to the absorption edge and limited birefringence of the crystal[9]. While the KBBF crystal allows implementation of high-resolution ARPES experiments with photon energies up to 7 eV[10-12], the momentum range covered by such light sources is still not wide enough to reach the edges or corners of the first Brillouin zone of typical solids (1 Å$^{-1}$). The following equation describes how much in-plane crystal momentum $k$ can be covered with a certain photon energy $\hbar\omega$ in a photoemission experiment:

$$k = \frac{\sqrt{2mE_{\text{kin}}}}{\hbar} \sin\theta$$

$$= \frac{1}{\hbar}\sqrt{2m(\hbar\omega - \varphi)} \sin\theta, \quad (1)$$

where $m$ is the free electron mass, $E_{\text{kin}}$ is the photoelectron kinetic energy, $\varphi$ is the work function of the solid, and $\theta$ is the polar emission angle with respect to the surface normal. It is clear from eq. (1) that higher photon energies yield larger photoelectron kinetic energies, and thus wider coverage of the momentum space.

High energy photons in the extreme ultraviolet (XUV) range are typically produced through high harmonic generation (HHG), which is a nonlinear optical process in which the frequency $\omega$ of an intense laser beam is up-converted to $n\omega$ through interaction with a medium (gas or solid, but typically inert gas). The output photon energies from HHG can range up to soft X-rays[13,14]. If inversion symmetry is present in such targets, only the odd-order harmonics are produced from the HHG process. Due to the extremely nonlinear nature of HHG, large electric fields are desirable to improve the generation efficiency. This requires using high laser pulse energies and shorter pulse durations as the seed pulse for HHG. As the pulse energy and the repetition rate are inversely proportional to each other for most laser systems (at constant average power), efficient HHG sources typically operate at low repetition rates (1-10 kHz). In this case, however, ARPES spectra often suffers from poor energy resolution due to the broad bandwidth of HHG pulses, and insufficient statistics as the number of electrons per pulse has to be limited in order to avoid severe space-charge effects[15-24]. Recent developments in fiber-based HHG setups, on the other hand, can often suffer from lattice heating effects due to the excessively high repetition rates[25,26]. It is therefore highly desirable to produce HHG pulses at an intermediate rate (100kHz – 1MHz) for time-resolved ARPES experiments in the weak perturbation regime.

In this work, we produce bright 11 eV laser pulses at high repetition rates (100 and 250 kHz) through a series of two harmonic generation stages – a third harmonic generation (THG, Light Conversion HIRO) of a 1038 nm, 190 fs Yb:KGW amplifier laser pulses (Light Conversion PHAROS) using nonlinear crystals, and another THG through a hollow-core fiber (KM Labs XUUS) filled with xenon gas. Since only 3 W of the amplifier output is routed toward the harmonic generation arm, the remaining 7 W is used to pump the optical parametric amplifier (OPA, Light Coversion ORPHEUS) that generates a tunable range of wavelengths from ultraviolet to near-infrared range. All elements of the system (OPA, DFG, and harmonic generator) are configured to be switchable between 100 and 250 kHz repetition rates through minimal optical alignment procedures.

**Results**

**Overview of the setup**

The overall scheme of the experimental setup is summarized in Fig. 1. The $3\omega$ beam is focused onto a ~100 μm spot (full-width at half-maximum, FWHM) at the entrance of the hollow-core fiber filled with xenon gas, in which another THG process up-converts part of the $3\omega$ beam to $9\omega$. We note that the fifth harmonic generation of $3\omega$ is extremely weak in the current configuration, indicating that the harmonic generation is highly perturbative. This process is clearly distinct from HHG, in which the flux of generated harmonics is similar in size over a wide range of harmonic orders ("plateau region")[13,14]. A set of waveplate and polarizer are also placed before the fiber to control the input power. Both the $3\omega$ and $9\omega$ beams co-propagate into the monochromator, which is directly connected to the hollow-core fiber under vacuum.

The monochromator mainly serves to filter out $3\omega$ and transmit $9\omega$ to the ARPES chamber. The off-plane Czerny-Turner monochromator was custom-built from McPherson Inc. (model: OP-XCT), and its design was inspired by the single-grating monochromator reported in Ref. 27. A more detailed description of the monochromator is also provided in Ref. 28. The gold-coated grating has 500 grooves per mm (Richardson gratings) and it is mounted in an "off-plane" geometry so that the rulings are oriented parallel to the propagation direction of the beam. Under such geometry, the following equation satisfies the diffraction condition[29]:

$$2 \sin\psi \sin\Lambda = n\lambda\sigma, \qquad (2)$$

where $\psi$ is the azimuthal angle of the grating, $\Lambda$ is the incidence angle of the beam with respect to the groove direction, $n$ is the diffraction order, $\lambda$ is the wavelength of the diffracted beam, and $\sigma$ is the groove density. The $9\omega$ (11 eV) beam travels at ~5° with respect to the grating, and a sharp diffraction peak is observed when $\psi$ is set to ~20°.

The output of the monochromator is terminated with a 0.5 mm thick LiF window (Korth Kristalle GmbH) mounted on an ultra-high vacuum (UHV) gate valve (VAT Inc.). During measurements, we use the LiF window to stop the inert gas flow to the UHV ARPES chamber. The typical transmission of 11 eV through the LiF window was measured to be ~20–30%. As the LiF window also transmits the $3\omega$ (3.6 eV) beam, the monochromator grating plays a crucial role in ensuring that only the $9\omega$ (10.8 eV) component is transmitted to the ARPES chamber. The 11 eV output beam then focuses at the exit slits of the monochromator. The slit openings can be tuned to control the spot size at the sample, as the exit slit is imaged to the sample in a 1:1 configuration.

The 11 eV beam then travels into the multi-port diagnostic chamber (Kimball Physics), in which the beam profile and intensity can be monitored. A toroidal gold-coated mirror can be inserted into the chamber to direct the 11 eV beam toward the CCD (Andor Newton), in which the overall intensity and fiber propagation mode of the 11 eV beam can be diagnosed from the CCD images. When the toroidal mirror is retracted from the beam path, the 11 eV flux can be measured more accurately with a photodiode (AXUV 100G, Opto Diode Corp.) by means of a lock-in detection (Table 1).

| Repetition rate | Input 3ω pulse energy | Photons/s |
|---|---|---|
| 100 kHz | 7 μJ | $4.7 \times 10^{10}$ |
| 250 kHz | 3 μJ | $2.4 \times 10^{10}$ |

**Table 1** | 11 eV output intensity measured at the diagnostic chamber (before the final focusing toroidal mirror)

We find that the 11 eV generation efficiency is higher at 100 kHz repetition rate under the same input power (pulse energy, on the other hand, is 2.5 times larger). This indicates that the THG process (3ω → 9ω) is highly nonlinear. The 11 eV beam then reflects off of a fixed toroidal mirror and focuses onto the sample mounted inside the ARPES chamber, in which the pressure is maintained at $< 1 \times 10^{-10}$ torr.

**Bi$_2$Se$_3$ measurements – time and energy resolution**

Assuming that various contributions to the broadening of spectra (11 eV bandwidth, analyzer resolution, etc.) can be collectively modelled with a Gaussian profile, the energy distribution curve (EDC) across the Fermi edge can be fitted using a Gaussian-convoluted Fermi-Dirac distribution:

$$I(E) = [f(E,T)D(E)] \otimes g(E)$$
$$= \frac{D(E)}{e^{(E-E_F)/k_B T}+1} \otimes e^{-4\ln 2\, E^2/\text{FWHM}_E^2}, \tag{3}$$

where $D(E)$ is the density of states, $E - E_F$ is the photoelectron energy with respect to the Fermi level, $k_B$ is the Boltzmann constant, $T$ is the sample temperature, and FWHM$_E$ is the energy resolution.

In order to evaluate the performance of our instrument, we acquired ARPES spectra close to the Fermi energy of an *n*-doped $Bi_2Se_3$ single crystal (see Methods). The Dirac cone surface states are clearly visible in Figs. 2a,b, in which cuts were taken along the orthogonal in-plane directions ($k_x$ and $k_y$). The Fermi energy is located 350~400 meV above the Dirac point, and signatures of the bulk conduction band can be observed close to the upper right region of the Dirac cone in Fig. 2a. By fitting the momentum-integrated EDC data to eq. (3), we obtain an energy resolution of 16 meV (Fig. 2c).

Since this monochromator was previously shown to improve the energy resolution of the ARPES setup based on HHG sources[28,30,31], it is important to establish whether the monochromator can cut down the 11 eV bandwidth any further. We thus measure the energy resolution of ARPES spectra taken at various slit openings (Fig. 3a). As shown in Fig. 3b, there is only a slight improvement in resolution at smaller slit openings. The poor energy resolution at the widest slit opening possibly originates from the instrument limitation as opposed to the broader bandwidth of the 11 eV source. This is because the calibrations of our time-of-flight electron analyzer are optimized for spot sizes of approximately 100 μm. Larger focal spot sizes generally increase uncertainties of the measured emission angle, and eventually broadens the measurements of photoelectron kinetic energies.

We now turn to time-resolved measurements of the setup. We choose 680 nm (1.82 eV) laser pulses from the OPA to pump the system to an excited state. The ARPES spectra are then taken at various time delays between the pump and probe pulses. The energy-momentum cuts in Fig. 4 shows that the upper region of the Dirac cone is populated at positive time delays after photoexcitation. The wide momentum coverage is apparent from the second panel, in which the photoexcited charge carriers populate

>0.7 eV above the Fermi level. While a 6 eV probe beam covers approximately ±0.19 A$^{-1}$ with ±15° angular range of the detector, our 11 eV setup can measure up to ±0.36 A$^{-1}$, which is almost four times as large as the momentum range of 6 eV measurements. The photoexcited charge carriers in Bi$_2$Se$_3$ gradually relaxes back to its equilibrium after a few picoseconds.

In order to characterize the time resolution of the setup, we plot the momentum-integrated photoelectron counts as a function of time delay at various energies above the Fermi level. The resulting contour (Fig. 5a) indicates that the relaxation dynamics of photoexcited charge carriers is faster for higher energies, typical of a metallic surface. The pump-probe traces are then plotted for different energies above the Fermi level by integrating the (momentum-integrated) photoemission counts over an energy window of 50 meV (Fig. 5b). These time-dependent photoemission counts can be fitted with a single exponential function convolved with a Gaussian function that takes the finite experimental time resolution into account:

$$I(t) = [Ae^{-\frac{t-t_0}{\tau}} \cdot H(t-t_0)] \otimes e^{-4\ln 2\, t^2/\text{FWHM}_t^2}, \qquad (4)$$

where $t - t_0$ is the pump-probe time delay, $\tau$ is the relaxation time of the pump-probe signal, $H(t - t_0)$ is the Heaviside step function, and $\text{FWHM}_t$ is the time resolution. For high energies above the Fermi level, the exponential decay is beyond the temporal resolution of the experiment. The time-resolved trace at this energy can then be described with a single Gaussian function. In fact, the traces that are too close to the Fermi level exhibit a slow rise time (possibly arising from the relaxation of higher energy electrons at ultrashort timescales), which obscures the proper determination of $\text{FWHM}_t$. The nonlinear least square fits of the data at various energies yield an upper bound time resolution of ~250 fs. The lower time resolution compared to the pulse width of the laser output (190 fs) probably originates from the optical elements inside the pump/probe paths, additional pulse broadening from the OPA unit, and a slightly non-collinear (~10°) propagation of the pump and probe beams.

**ErTe$_3$ measurements – recovery dynamics of two CDW regions**

To further demonstrate the capability of the 11 eV setup, we present tr-ARPES spectra of a rare-earth tri-chalcogenide system ErTe$_3$. The family of $R$Te$_3$ compounds ($R$ = rare earth element) forms a layered orthorhombic crystal structure, and typically exhibits charge density wave (CDW) gaps across a large region of the Fermi surface[32]. Prior tr-ARPES measurements on TbTe$_3$ identified amplitude modes[6] and demonstrated that the CDW state may be transiently stabilized upon near-infrared excitation[33]. A previous work using this 11 eV setup also identified a photoinduced CDW phase transition whose recovery dynamics is mediated through topological defects[34].

Among the $R$Te$_3$ compounds, the heavier ErTe$_3$ exhibits two separate CDW phase transitions (at 155 K and 267 K) with ordering wavevectors oriented along the two orthogonal in-plane directions (see Fig. 6a)[35,36]. The dual-CDW compound ErTe$_3$ provides an ideal platform to use the 11 eV system, as the momentum coverage is sufficiently wide to cover both CDW regions simultaneously.

Figure 6a shows a constant energy contour at the Fermi level, in which a large fraction of the ErTe$_3$ Fermi surface is gapped out due to the CDW formation (Fig 6a). By taking energy-momentum cuts (Figs. 6c,d) across the gapped regions (green horizontal dashed lines in Fig. 6a), the leading edge of the valence band yields the CDW gap sizes of ~250 meV (CDW1) and ~50 meV (CDW2), which are consistent with another ARPES measurement[36]. The distinct separation between the two CDW regions is apparent in Fig. 6b, which shows a cut taken along the vertical blue dashed line shown in Fig. 6a.

We then performed time-resolved measurements using 720 nm (1.72 eV) pump pulses, whose photon energy exceeds the size of both CDW gaps. The snapshots of energy momentum cuts (Fig. 7) reveal that the CDW gaps are transiently filled right after photoexcitation, similar to the observations reported in Refs. 6,33, suggesting a transient suppression of the CDW orders and a photoinduced phase transition to the normal state. The extent of gap filling is maximized at ~250 fs after the pump pulse arrival, indicating that ultrafast photoexcitation across the gap (Figs. 7a,b) precedes the photoinduced phase transition. Within 2 ps however, the population of photoexcited carriers above the Fermi level and within the gap almost completely vanishes, and the electronic structure recovers to a quasi-equilibrium state. A key strength of our setup is the ability to simultaneously monitor the dynamics of two CDW orders within a single measurement. A close look at the relaxation dynamics within the gap reveals that the recovery is much slower for the case of CDW2 (relaxation times: $\tau_1 = 2.4 \pm 0.3$ ps and $\tau_2 = 0.60 \pm 0.07$ ps), whose gap size is significantly smaller. This observation indicates that the ordered bound-state electrons under the larger gap ($\Delta_{CDW1}$) have a stronger tendency to form a CDW, and consequently the relaxation is much faster than in the case of the smaller gap ($\Delta_{CDW2}$). Another possibility is a larger phase space available for the charge carriers relaxing across $\Delta_{CDW1}$, leading to a faster recovery time. The detailed physical mechanism leading to different dynamics in the two gaps will be presented elsewhere.

**Discussion**

The wide momentum coverage of 11 eV, combined with the 3D mapping capability of the time-of-flight electron analyzer provides a powerful method of studying the nonequilibrium band structure on sub-picosecond timescales. The tunable, fast repetition rates (100/250 kHz) of the system allows high acquisition rates as the photoelectron pulses undergo less space-charge repulsions. The tunability of pump photon energies also allows the system to be photoexcited under various resonance/off-resonance conditions with minimal adjustments. Above all, the unprecedented energy and time resolutions of 16 meV and 250 fs with XUV pulses enable a high-resolution, momentum-selective non-equilibrium studies of band structure across a wide range of the Brillouin zone.

*Note*: During the preparation of this manuscript, we became aware of two related works, in which the authors use the second harmonic of a YB:KGW amplifier output as the seed pulse[37,38]. We note that the pulse energies used in these setups are significantly higher (>70 µJ) than those used in this work (3-7 µJ).

**Acknowledgements**

We thank Ya-qing Bie and Pablo Jarillo-Herrero for assistance with the ErTe$_3$ sample preparation using the glove box, and Fahad Mahmood for useful discussions during the early stage of this work. This work was supported by the U.S. Department of Energy, BES DMSE grant number DE-FG02-08ER46521 (planning, instrumentation and data taking), Army Research office (support for time of flight detector) and by the Gordon and Betty Moore Foundation's EPiQS Initiative grant GBMF4540 (support for the laser). Crystal growth and characterization of the ErTe$_3$ samples were performed at Stanford University with support from the Department of Energy, Office of Basic Energy Sciences under contract DE-AC02-76SF00515.## Methods

### Bi$_2$Se$_3$ measurements

The Bi$_2$Se$_3$ sample was cleaved at 35 K under a base pressure of $1.0 \times 10^{-10}$ torr. For static measurements, the time-of-flight (ARTOF 10K from Scienta Omicron) detector was configured with a ±15° emission angle range and a 10% energy window under a double-pass mode. For time-resolved measurements, the angular range was set to ±15° and the energy window was set to 20% under a double-pass mode. All measurements were taken at 250 kHz.

### ErTe$_3$ measurements

Single crystals of ErTe$_3$ grown by slow cooling a binary melt[35] are air sensitive. Therefore, all procedures of sample preparation are carried out in a glove box before the sample is mounted in the ARPES chamber. After minimal exposure to ambient pressure, the samples were cleaved *in situ* under a base pressure of $1 \times 10^{-10}$ torr at 15 K ($< T_{\text{CDW1}}$ and $T_{\text{CDW2}}$). All measurements were taken at 100 kHz.

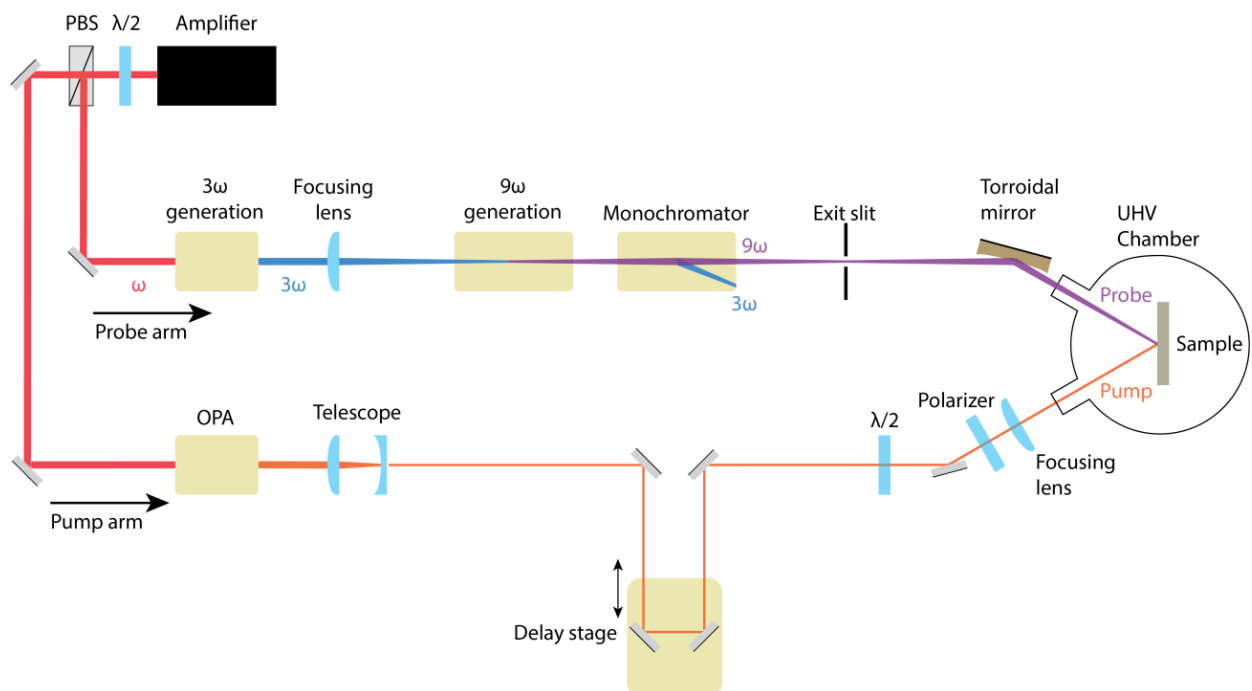

**Figure 1 | The 11 eV setup.** A simplified diagram of the beam paths in the 11 eV tr-ARPES setup. OPA: optical parametric amplifier, PBS: polarizing beam splitter.

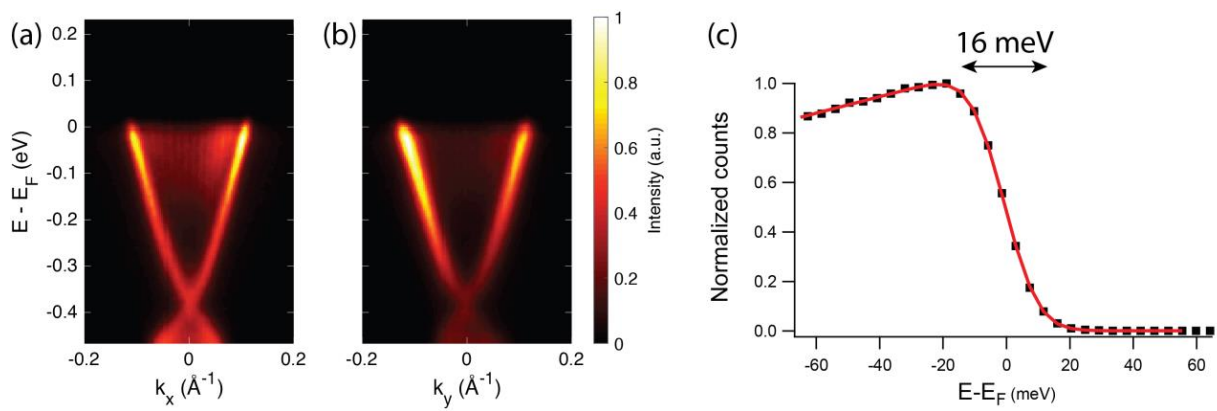

**Figure 2 | ARPES spectra from $Bi_2Se_3$ single crystals.** The linearly dispersing surface states are clearly visible in cuts along the **a,** $k_x$ and **b,** $k_y$ directions. **c,** A momentum integrated (across ±0.2 Å$^{-1}$ along both $k_x$ and $k_y$ directions) EDC yields a 16 meV energy resolution.

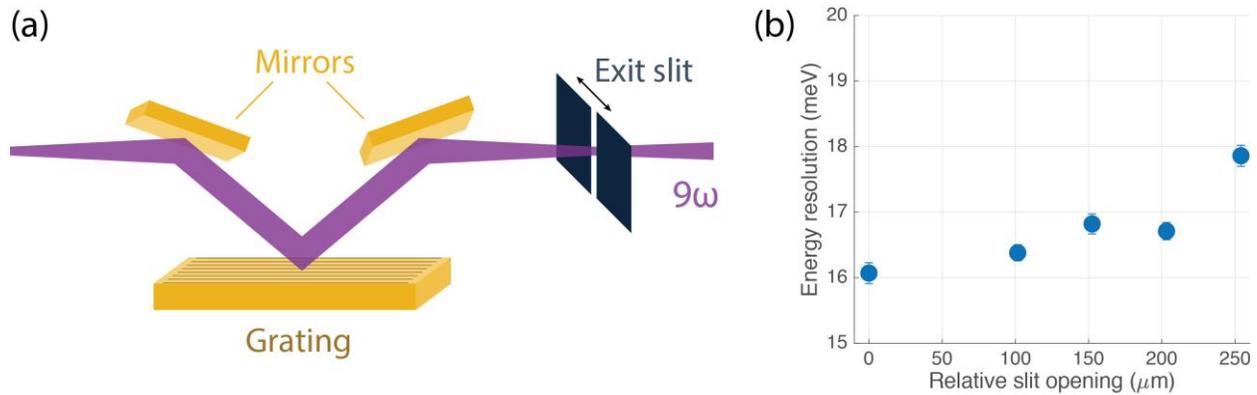

**Figure 3 | Slit dependence of the energy resolution. a,** A schematic showing the beam path through the grating and the exit slit. ARPES spectra were taken at various horizontal slit opening values. **b,** Energy resolution remains almost unchanged at different slit settings. Error bars represent the uncertainties of least square fitting of momentum-integrated (across ±0.2 Å$^{-1}$) EDC data to eq. (3).

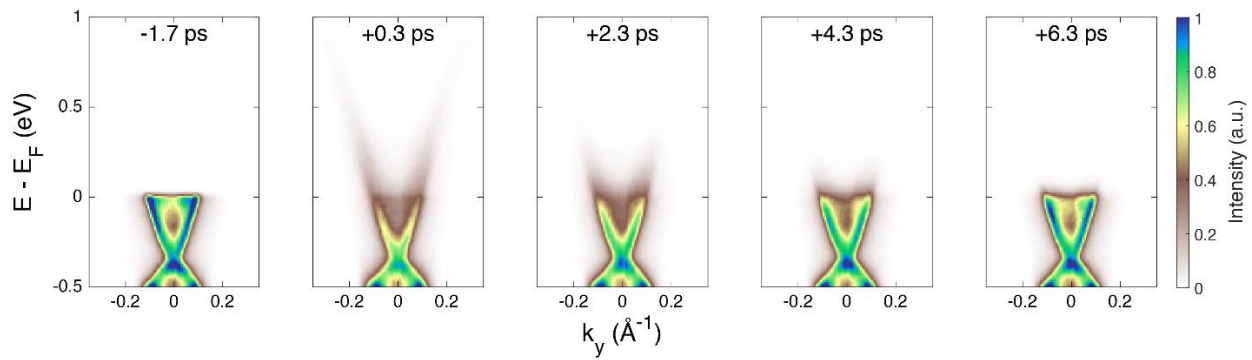

**Figure 4 | Snapshots of Bi$_2$Se$_3$ tr-ARPES spectra.** Energy-momentum cuts were taken across the Γ point at different time delays (displayed in the upper part of each panel) between the pump and probe pulses. The upper part of the Dirac cone is transiently populated due to photoexcitation by the 1.82 eV pump pulse.

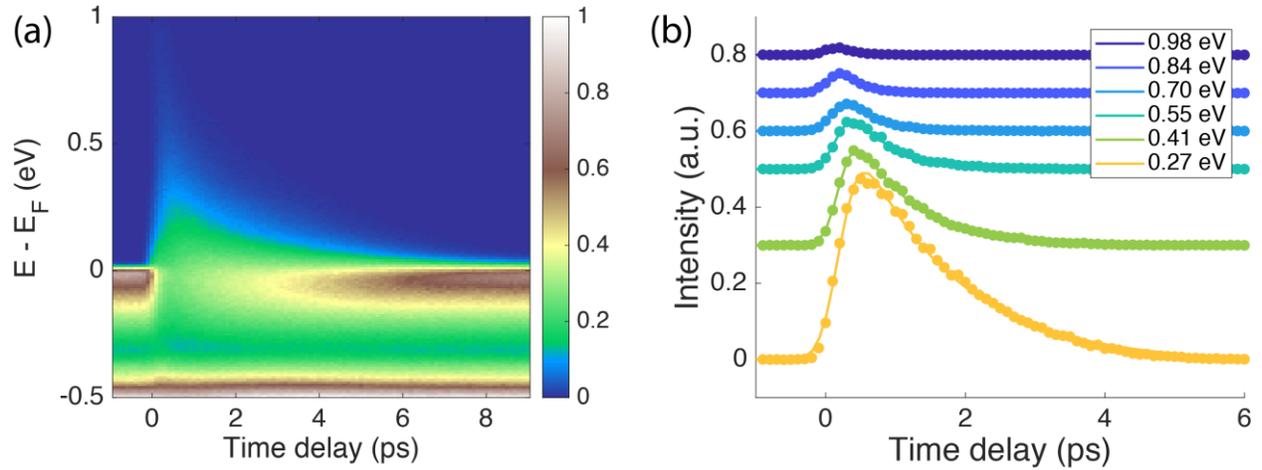

**Figure 5 | Momentum-integrated, time-resolved photoemission traces of Bi$_2$Se$_3$. a,** A two-dimensional plot of the momentum-integrated photoemission (counts at different time delays and energies with respect to the Fermi level. **b,** A plot of photoemission counts as a function of time delay at different energies relative to the Fermi level. Each trace was integrated over a 50 meV window. Raw data (filled circles) and fits (curves) to eq. (4) are in good agreement.

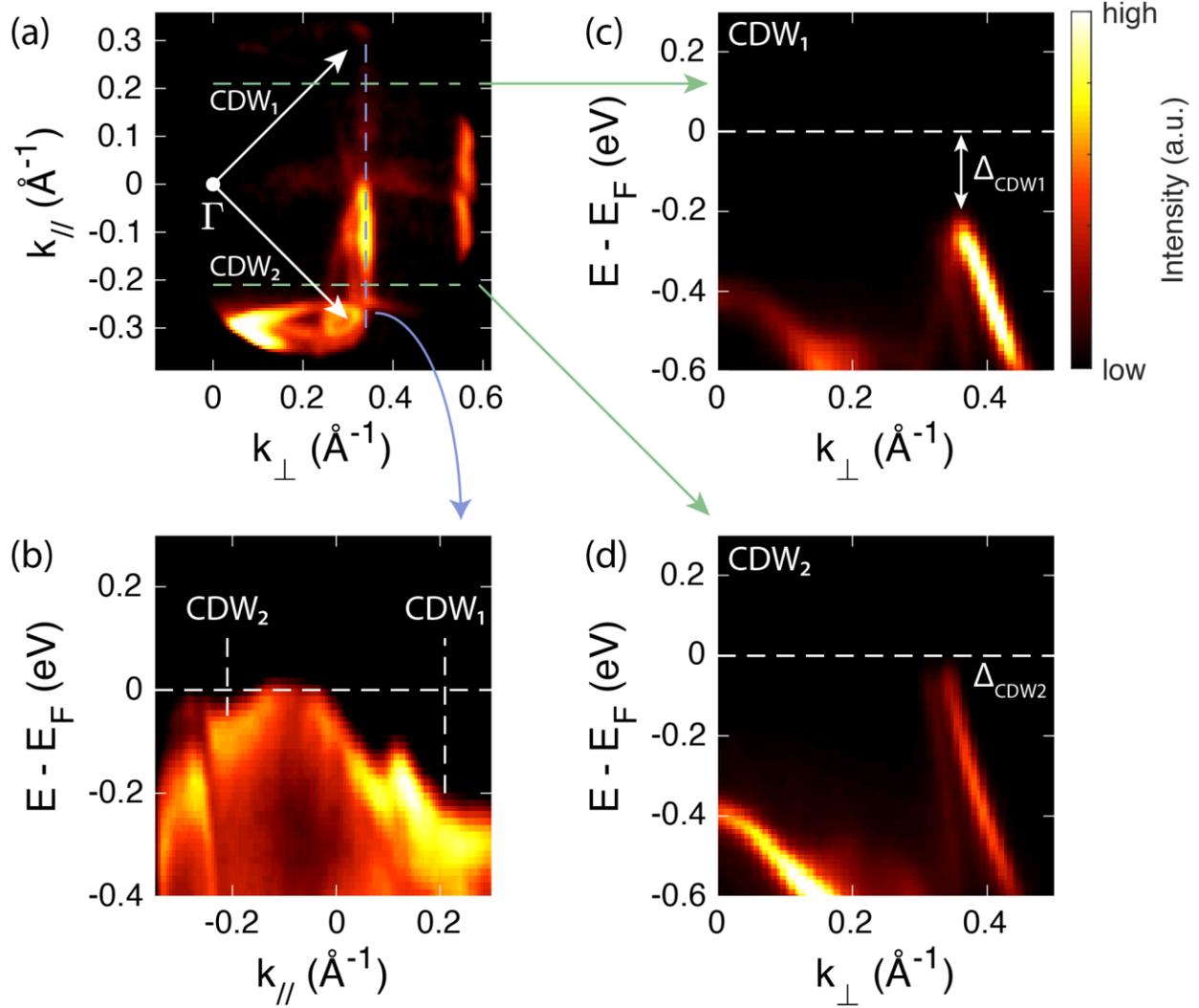

**Figure 6 | Static ARPES spectra of ErTe3. a,** The Fermi surface (integrated over 50 meV) exhibits a large gapped area due to the CDW formations. Energy momentum cuts were taken across the **b,** vertical (blue) and **c,d** horizontal (green) lines displayed in **(a)** with an integration window of 0.02 Å⁻¹. Two distinct regions of CDW gaps are clearly observable. The momenta $k_{//}$ and $k_\perp$ are oriented at 45° with respect to the crystallographic axes *a* and *c* along the basal plane.

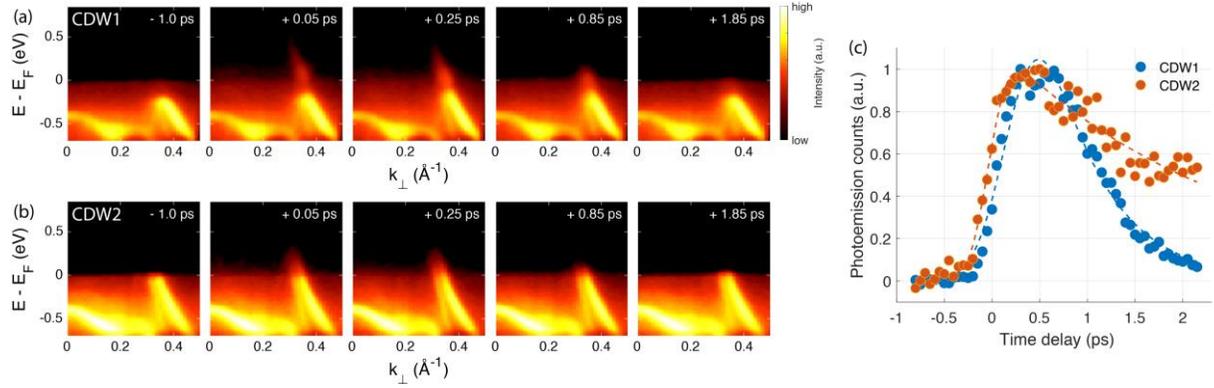

**Figure 7 | tr-ARPES snapshots of ErTe$_3$.** Energy momentum cuts are taken across the **a,** CDW$_1$ and **b,** CDW$_2$ gaps (green dashed lines shown in Fig. 6a) at different pump and probe time delays. The CDW gap closings are observed 0.25 ps after the pump pulse arrival. **c,** The two CDW regions show different relaxation timescales upon pump excitation. The photoemission counts were integrated across momentum and energy ranges of ±0.05 Å$^{-1}$ and ±25 meV, respectively around the gapped regions. The resulting data (dots) were then fit (dashed curves) to eq. (4) in order to extract the relaxation times.